# Efficient Light Trapping Strategy in Solar Cells via Vertically Disordered Perovskite Nanowires

Sina Salimian and Nicklas Anttu

Physics, Faculty of Science and Engineering, Åbo Akademi University, FI-20500 Turku, Finland

E-mail: sina.salimian@abo.fi

**Abstract**

In this work, we investigated the absorption of light in perovskite nanowires (PrvNWs) for application as active layer in a solar cell. Specifically, we simulated PrvNWs with a random position inside the pores of an AAO membrane. The results demonstrate that this vertical disorder can boost light absorption, up to 28% in terms of short-circuit current, in comparison to the corresponding ordered PrvNW array. This performance boost is due to a combination of light trapping and anti-reflection effect. Thus, our findings suggest that vertically disordered PrvNWs can have a much better performance than ordered arrays. Also, we showed that it is not necessary to aim for a very narrow length distribution in the nanowires, easing the fabrication requirements.

**Introduction**

Metal halide perovskites (MHPs) have become a well-established class of compounds integrated in perovskite solar cells (PSCs), but there still remains much to discover about their true potential [1,2]. MHPs are chemically formalized as $ABX_3$ where A, B, and X are organic/inorganic cations, metal cations, and halide anions, respectively [3]. For example methylammonium lead halide MHP has been used successfully in PSCs [4]. The reason for the interest in MHPs lies in their materials properties and fabrication cost. Their optical and electrical features are promising, including bandgap tunability, large absorption coefficient, excellent charge-carrier mobilities, and long minority carrier diffusion length. Additionally, solution-based synthesis methods make MHP-based optoelectronic devices much more affordable to fabricate compared to those based typical inorganic semiconductors like Si and especially III-V semiconductors like GaAs and InP [5,6,7]. Demonstration of the performance of MHPs is seen for example in the latest efficiency report of 35.2% for a hybrid tandem MHP/Si solar cells [8]. Thus, their materials properties and cost-effectiveness make MHPs highly promising candidates for the next generation of practical solar cells.

Commercialization of PSCs, however, has been stalling especially due to their limited long-term stability despite their promising performance [9,10]. MHPs are known to be sensitive to environmental elements such as oxygen, heat, and moisture. Over time, this sensitivity may lead for example to the formation of lead halide compounds which have negative impact on the performance of the PSC [11]. In addition, formation of non-uniform regions in the MHP over time could result in degrading charge extraction from the MHP to the electrodes [12].

One promising way for enhancing the stability of PSCs could be the usage of MHPs in the form of nanowires (NWs), for example fabricated in an ordered array of pores in an oxide membrane such as anodized aluminum oxide (AAO) [13]. Such MHP NWs (PrvNWs) fabricated in an AAO membrane were found highly stable [13]. There, degradation is restricted to each individual PrvNW, instead of permitting widespread damage throughout the entire sample as in a bulk MHP layer [14]. Also, the PrvNW geometry could offer better mechanical tolerance to stress applied to the sample, allowing for highly flexible solar cells [15]. For electronic properties, the PrvNW geometry could allow for synthesis with fewer defects such as grain boundaries than in a bulk MHP, resulting in a decrease of unwanted non-radiative charge carrier recombination [16,17].

Due to the sub-wavelength size of PrvNWs, diffraction effects determine the absorption of light. Thus, the absorption in PrvNWs is affected by the geometry and the refractive indexes of the constituent materials, that is, both the MHP material and the surrounding materials [18]. For example, in an ordered array of PrvNWs, with optimized diameter ($D$) for the PrvNWs and array period ($P$), for a PrvNW length ($L$) of 1500 nm, the absorption is so strong that the resulting short-circuit current ($J_{sc}$) exceeds 90% of the $J_{sc,max}$ that occurs when all above bandgap photons of sunlight are absorbed [19].

NW arrays are promising structures for enhancing absorption through light-trapping in varying randomized configurations. For SiNWs, by introducing random in-plane position for the NWs in the array, stronger absorption was obtained in comparison to its corresponding ordered, non-random array counterpart [20]. Similarly, an ordered array of SiNWs but with random diameter and length for each NW increased the absorption [21].

In the case of PrvNWs fabricated in the pores of an AAO membrane, a new type of randomness is enabled by random vertical position of each NW, that is, out-of-plane randomness [22]. Such randomness might occur naturally in fabrication [23]. If such randomness can be tolerated, manufacturing costs might be reduced thanks to decreasing the need for strict precursor concentration control, the reduction of the precise growth-time of NWs, and fewer optimization iteration cycles to ensure the uniformity of the array [24,25]. Promisingly, initial studies have indicated potential for inducing additional light-trapping inside the membrane with such randomness [26]. However, optics modelling studies to explain and optimize that light-trapping are needed.

Therefore, in this paper, we numerically studied and modeled the light absorption in a vertically disordered array of PrvNWs using the finite-difference time-domain (FDTD) method. First, we simulated the ordered PrvNWs (see Fig. 1a for schematic) with optimized diameter and period in an AAO membrane and used the extracted absorption curves as reference for assessing effects from light-trapping in out-of-plane randomized structures (see Fig. 1b for schematic). Second, we studied the effect of light-trapping by increasing the thickness of the AAO membrane while keeping the PrvNW dimensions constant, while giving the PrvNWs a random vertical distribution within the AAO membrane. Third, we simulated the disordered PrvNWs with varying refractive index $n_{pore}$ at the top and bottom side of the PrvNWs to account for the case of electron and hole transport layer (ETL/HTL) material inserted into the pores to contact the NWs. Lastly, we considered also a random length distribution for the PrvNWs to cover the situation in which PrvNWs may grow in different length inside the porous AAO membrane.

Our results show that in general, there is not a strict requirement on planarizing the PrvNW layer to optimize the absorption. Thus, the synthesis with variation in vertical nanowire position can be allowed for, without drop in absorption.

Even more interestingly, we show that especially for ETL/HTL materials of low refractive index, we find strong light-trapping, further aided by an anti-reflection effect, that boosts the absorption. The boost is by up to 26%, in terms of short-circuit current, in our study. Furthermore, we find that random length of the disordered nanowires does not limit the light-trapping. Thus, our results suggest that inherently random growth of perovskite nanowires – often portrayed as a challenge when aiming to fabricate uniform arrays – can be beneficial for the absorption of light.

**Methods**

We simulate the optical response of vertically disordered PrvNWs using the FDTD method for two representative MHPs: $CH_3NH_3PbI_3$ (MAPI) and $CsPbBr_3$ (CPB), with rather different bandgap energy of approximately 1.57 eV and 2.33 eV, respectively. Their respective refractive indexes $n$ were extracted from Refs. [22,27] (Fig. S1).

To make the study relevant for PSCs, we consider the following short-circuit current, where we assume that each absorbed photon gives rise to a charge carrier in the short-circuit current:

$$J_{sc,A} = q \int_{\lambda_{short}}^{\lambda_{bg}} \frac{I_{AM1.5}(\lambda)A(\lambda)}{2\pi\hbar c/\lambda} d\lambda. \quad (1)$$

Here, $A$ is the FDTD-calculated absorption spectrum of the PrvNWs, $q$ the elementary charge, $\hbar$ the reduced Plank constant, $c$ the speed of light in vacuum, $I_{AM1.5}$ the AM1.5G incident solar spectrum [28], $\lambda_{bg}$ the bandgap wavelength of the PrvNWs (~790 nm and ~530 nm for MAPI and CPB, respectively), and $\lambda_{short}$ of 300 nm the lower limit below which the solar intensity is negligible. To evaluate the optical absorption of the PrvNWs, we calculated the overall light absorption by two monitors, one for reflection $R$ at the top and the other for transmission $T$ at the bottom of the membrane, yielding $A = 1 - R - T$. To analyze reflection and transmission losses, we use $R$ or $T$, respectively, in Eq. (1) instead of $A$ to yield the corresponding loss $J_{sc,R}$ and $J_{sc,T}$. The maximum possible short-circuit current, $J_{sc,max}$, is obtained when $A = 1$ is used in Eq. (1) for all above bandgap photons. Then, $J_{sc,A} = J_{sc,max} - J_{sc,R} - J_{sc,T}$.

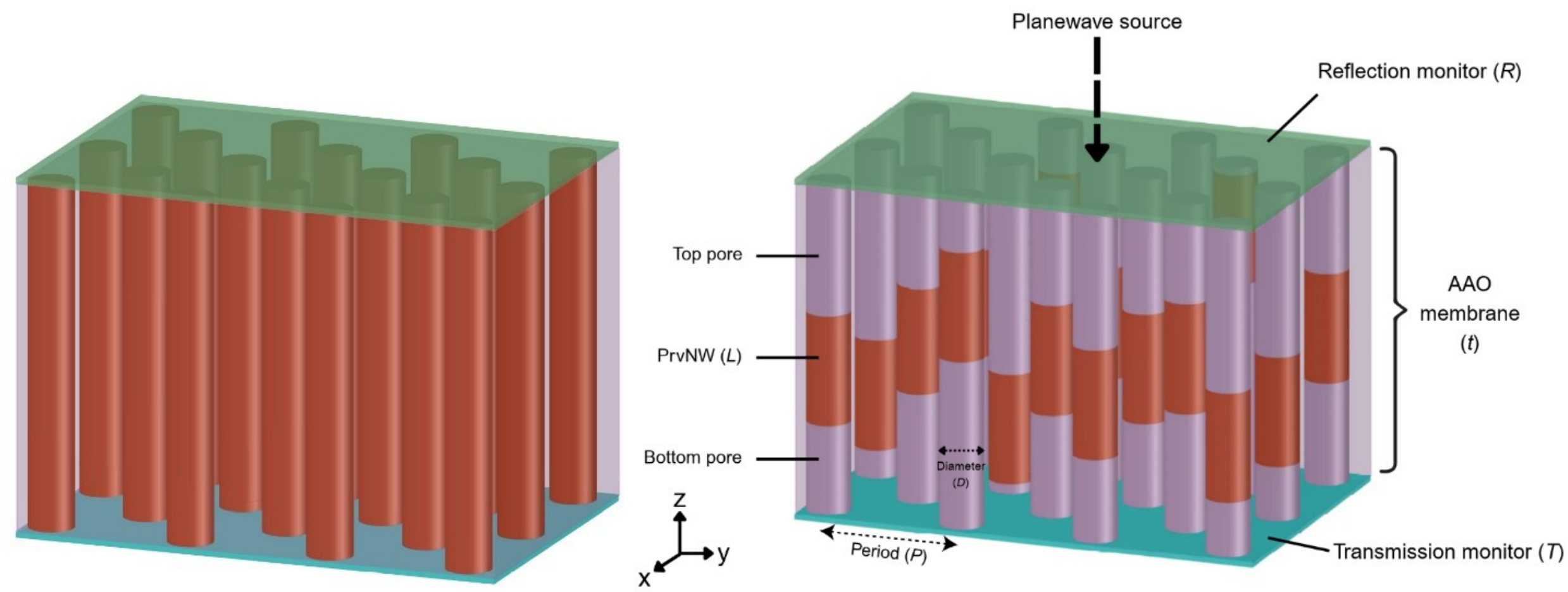

Fig. 1 PrvNWs embedded in a porous AAO membrane in (a) ordered and (b) vertically disordered fashion. As it is shown in (b), each PrvNW has top/bottom pores to account for the ETL/HTL material inside the pores. The monitor planes to extract reflection and transmission spectra are indicated.

For reference, we simulated the absorption performance and corresponding $J_{sc}$ for ordered arrays of vertically aligned PrvNWs embedded in an AAO membrane (Fig. 1(a)). For both MAPI and CPB PrvNWs, we calculated the optimum diameter ($D_O$) and array period ($P_O$) with the methodology described in our previous work [19]. This optimization was performed for PrvNWs with length of $L$ = 500 nm, which is our main focus for MAPI PrvNWs below. The extracted $D_O$ and $P_O$ were 380 nm and 440 nm for MAPI, respectively, while for CPB, they were 370 nm and 510 nm (see Figure 2 for the resulting absorption spectra).

We introduced disorder into the system by vertically displacing the PrvNWs within an AAO membrane with different thicknesses $t$ (Fig. 1(b)). The vertical position of each PrvNW was randomly set from a uniform $z$ distribution in such a way that neither the top nor the bottom of the PrvNW stick out from the AAO membrane.

In these simulations, the membrane thickness $t$ was varied, and we considered PrvNWs with $L$ = 500 and 1000 nm, with material of $n_{pore}$ (initially set to 1) filling the pores on the top and at the bottom of PrvNWs. For nanowires with $L$ = 500 nm, $t$ was varied from 600 nm to 1000 nm with 100 nm steps, followed by the thickness of 1500 nm, and then from 2000 nm to 5000 nm with 1000 nm step. For $L$ = 1000 nm, $t$ ranged from 1100 nm to 1500 nm in 100 nm steps and from 2000 nm to 5000 nm with 1000 nm steps.

For each set of such geometry parameters, we simulated four realizations of the quasi-random PrvNW system. For all the arrays that we studied, the specific realization of the quasi-random configuration showed only minor effect on the resulting optical response (as seen by the variation of the short-circuit extracted from each set of four simulations at a single thickness in Fig. 3 and 4, marked by the error bars, and from Fig. S2).

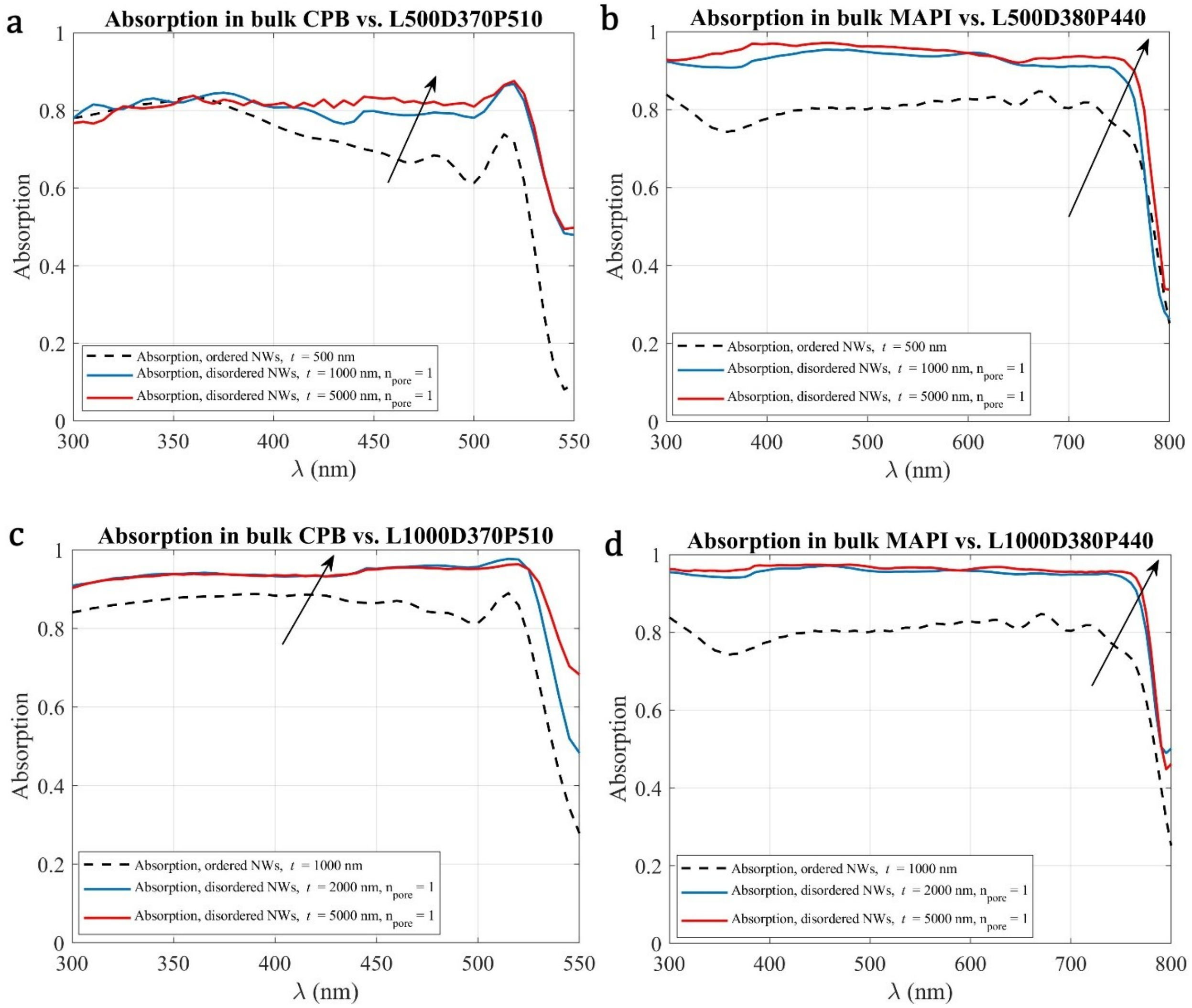

Fig. 2. The absorption of light in a periodic hexagonal array of (a) MAPI PrvNWs and (b) CPB PrvNWs of $L$ = 500 nm in (a) and (c); and $L$ = 1000 nm in (c) and d), embedded in AAO membrane. For disordered NWs, two different thicknesses of $t$ = 2000 nm (blue lines) and 5000 nm (red lines) were simulated to verify their performance in comparison to their relative ordered version (dashed lines). Here, the absorption curve for one quasi-random realization is shown.

## Results and Discussion

In Fig. 2, we compare for $L$ = 500 and 1000 nm an ordered PrvNW array with $t = L$ and disordered PrvNW arrays with $t$ = 1000 and 5000 nm; and $n_{pore}$ = 1. The increase in $t$ for the disordered arrays leads to noticeably increased absorption for both MAPI and CPB PrvNWs. In the case of CPB, a clear optical saturation for $L$ = 500 nm at 300 nm < $\lambda$ < 360 nm is visible in Fig. 2b preventing any increase of absorption in that wavelength range. The 1000 nm-long CPB PrvNWs, however, allow for higher absorption for larger $t$ for all the wavelengths studied.

The increasing absorption with increasing membrane thickness could originate from light scattering from the disordered PrvNWs, enhancing the optical path length inside the array as well as allowing excitation of varying types of optical eigenmodes in the PrvNWs at high internal angles, as compared to the ordered array [29]. We assume that such light-trapping is central for the better performance of disordered PrvNWs against the ordered arrays, allowing the PrvNWs to move toward the maximum absorption in Fig. 2.

Next, we study how changing the refractive index of the pores above and below the PrvNWs affects the absorption of light in the PrvNWs. Such a study is relevant in the context of an active PrvNW layer in a solar cell configuration, for which we assume electron and hole transport layer (ETL and HTL) materials, represented optically by $n_{pore}$, inside the pores to connect the PrvNWs electrically. At the same time, by varying $n_{pore}$, we aim to determine whether the increased absorption in the membrane is caused mainly by light trapping via vertically disordered PrvNWs, or if it is driven by an anti-reflection effect. In a system with enhanced absorption, (i) decreased transmission without decrease in reflection indicates light-trapping dominated absorption, (ii) lower reflection combined with higher transmission suggests anti-reflection effect without light-trapping, and (iii) simultaneously lower reflection and lower transmission indicates a combination of light-trapping and anti-reflection effect.

We studied $n_{pore}$ with values of 1, 1.3, 1.5, $n_{AAO} \approx 1.78$, 2, and 2.5, extracting $J_{sc,A}$, $J_{sc,R}$, and $J_{sc,T}$, for the values of $t$ considered. We focus on MAPI PrvNWs with $L$ = 500 nm and CPB PrvNWs with $L$ = 1000 nm. The reason for this difference in $L$ for MAPI and CPB is the difference in their absorption coefficient [which is lower for CPB [22], (Fig. S1)—hence, we compensate by a larger $L$ for CPB here].

At $t$ = 600 nm, the $J_{sc,A}$ of the MAPI PrvNWs spanned from 20.5 mA/cm$^2$ to 22.5 mA/cm$^2$ for the varying $n_{pore}$ (Fig. 3a), as compared to 19.4 mA/cm$^2$ for the ordered array (dashed line in Fig. 3a). For $t \geq 700$ nm, the lower the $n_{pore}$, the higher the $J_{sc,A}$. At $t$ = 2000 nm, the difference between $n_{pore}$ = 1 and 2.5 is 5.3 mA/cm$^2$ in $J_{sc,A}$, and for $n_{pore}$ = 1, we have an increase by 28% compared to the case of the ordered array, showing the effectiveness of the randomized vertical positioning of NWs for closing the gap between $J_{sc,A}$ and the $J_{sc,max}$ = 26.5 mA/cm$^2$ for MAPI (dot-dashed line in Fig. 4a).

The CPB PrvNWs (Fig. 3b) follow a similar same trend as MAPI PrvNWs of decreasing $J_{sc,A}$ with the increase of $n_{pore}$, with 7% increase of $J_{sc,A}$ for $t$ = 2000 nm at $n_{pore}$ = 1. We note for both MAPI and CPB PrvNWs (Fig. 3a and 3b) that for the lowest $n_{pore}$, we find enhanced absorption for all values of $t$ studied. However, for increasing $n_{pore}$, we find that with increasing $t$, $J_{sc,A}$ can actually drop below the value of the ordered PrvNW array. Thus, there is strong dependence on the optical response on $n_{pore}$ and $t$.

To explore the origin of the above behavior in $J_{sc,A}$, we turn to $J_{sc,R}$ and $J_{sc,T}$ (Fig. 3c-f) since $J_{sc,A} = J_{sc,max} - J_{sc,R} - J_{sc,T}$. The vertical disorder caused the reflection to drop below that of the ordered counterpart (except for a minor increase in $J_{sc,R}$ for $n_{pore} = n_{AAO}$ for the CPB PrvNWs). At the same time, we find lower $J_{sc,T}$ at most thicknesses with $1 \leq n_{pore} \leq n_{AAO}$. Thus, there, we assign the enhancement of $J_{sc,A}$ to a combination of effective light-trapping and anti-reflection effect. With the increase of $n_{pore}$, however, the light transmission through the membrane increases for higher thicknesses, reaching values above that of the ordered array (Figs. 3e and 3f). There, a possible increase in $J_{sc,A}$ is instead assigned solely to an anti-reflection effect.

For the MAPI PrvNWs, for $n_{pore}$ = 1, 1.3 and 1.5, both $J_{sc,R}$ and $J_{sc,T}$ are considerably lower than in the ordered array, with up to 3.34 mA/cm$^2$ lower reflection loss and up to 1.97 mA/cm$^2$ lower transmission loss at $t$ = 1500 nm for $n_{pore}$ = 1. These results thus suggest strong light-trapping for vertically disordered MAPI PrvNWs with as low $n_{ETL/HTL}$ as possible.

For CPB PrvNWs, the $J_{sc,A}$ for the ordered array outperforms that of the disordered array with $n_{pore}$ = $n_{AAO}$, 2, and 2.5 already at smaller values for $t$ than for the MAPI PrvNWs. This originates from the higher $J_{sc,R}$ of $n_{pore}$ = 2.5 (Fig. 3d) and $J_{sc,T}$ of $n_{pore}$ = $n_{AAO}$, 2, and 2.5 (Fig. 3f) for the CPB PrvNWs. For $n_{pore}$ = 1 and 2.5, $J_{sc,T}$ is considerably lower in the CPB PrvNWs than in the ordered array, thus indicating light-trapping.

For neither CPB nor MAPI PrvNWs do we find for any $n_{\mathrm{pore}}$ considered a noticeable increase in $J_{sc,T}$, compared to the ordered counterpart, for $t \leq 1000$ nm for the MAPI PrvNWs and for $t \leq 1500$ nm for CPB PrvNWs. Thus, for this range of $t$ and $n_{pore}$ combinations, there is not a need to planarize the PrvNW layer in order to avoid light leaking through the disordered membrane, thus lifting the requirement of planarization in fabrication.

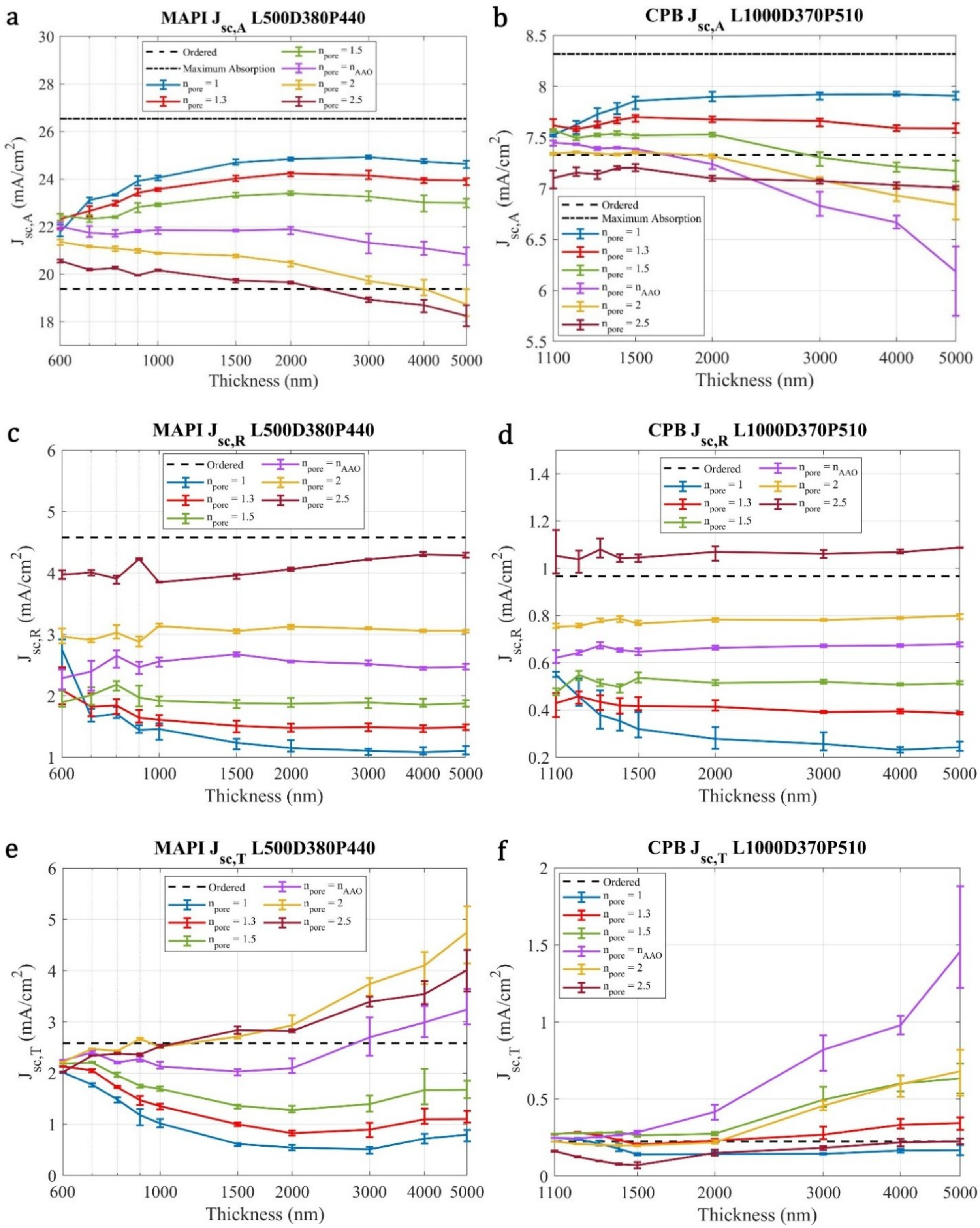

Fig. 3 Optcal response in terms of short-circuit current from Eq. (1) for disordered PrvNWs. (a), (c), and (e) refer to $J_{sc,A}$, $J_{sc,R}$, and $J_{sc,T}$, of MAPI. (b), (d), and (f) refer to $J_{sc,A}$, $J_{sc,R}$, and $J_{sc,T}$, of CPB NWs. The errorbars show the range of values from the four quasi-random arrays that were simulated. Dashed lines show the values for the corresponding ordered array, and dashed-dotted line in (a) and (b) show $J_{sc,max}$.

Finally, we explored the effect of introducing randomness in PrvNW length in the AAO membrane. We simulated membranes with out-of-plane PrvNWs with random length of 100 nm < $L < t$ within the membrane of $t$ in order to investigate the light behavior in a membrane with uneven length-growth of PrvNWs (see Fig. 4a for schematic).

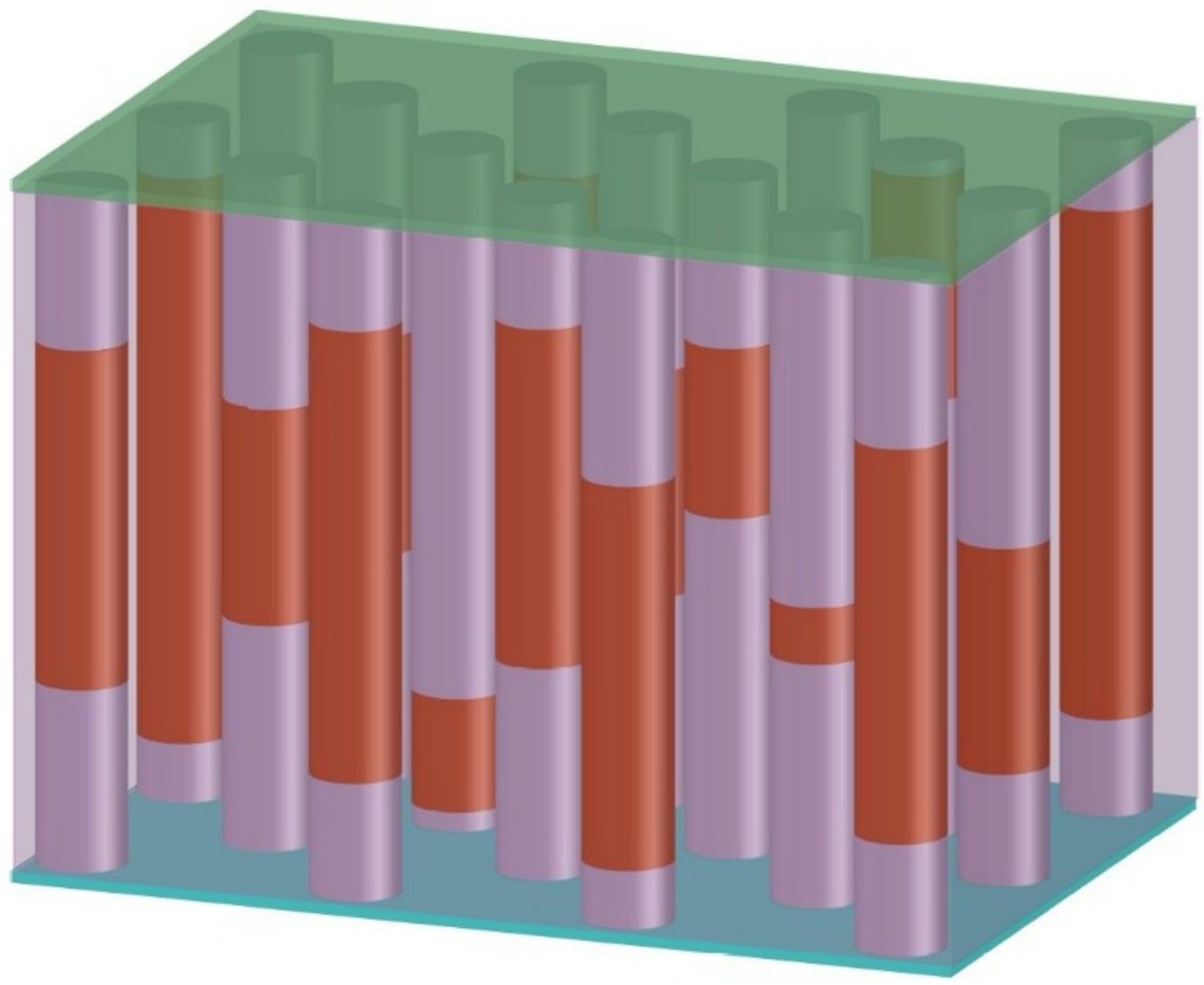

Fig. 4 PrvNWs with varying length embedded inside a porous AAO membrane.

We compared the absorption of these random arrays against ordered PrvNW arrays with the PrvNWs spanning the full thickness of the membrane, that is, with $L = t$. The initial assumption was that the ordered version with $L = t$ would have a better optical performance in comparison to the disordered PrvNWs with $L < t$ due to the larger volume of perovskite material in the ordered version. As can be seen in Fig. 5, however, randomization of length has a strong impact on $J_{sc,A}$ at low $n_{pore}$, especially for MAPI, with nearly 16% increase for $t = 2000$ nm. When considering $J_{sc,R}$ and $J_{sc,T}$ (Fig. S3), we can assign this increase of absorption predominantly to the anti-reflection effect rather than to light trapping, unlike for the case of PrvNWs with fixed length in Fig. 3. From these results, we conclude that it is not necessary to strive for a very narrow length distribution in the PrvNW fabrication.

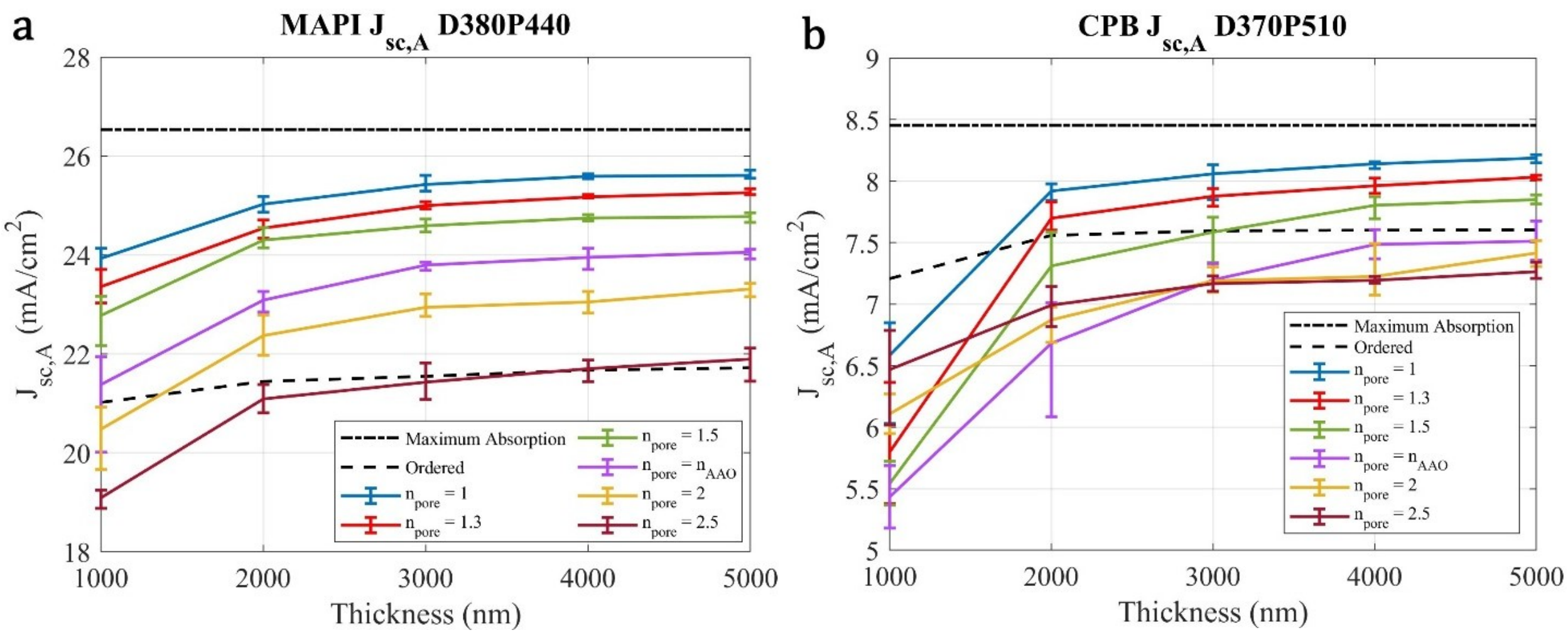

Fig. 5 The calculated short circuit current for PrvNWs with different lengths of $L < t$ (solid lines) for (a) MAPI, and (b) CPB PrvNWs. The dashed lines are for the ordered version of PrvNWs with $L = t$, and the dot-dashed lines show $J_{sc,max}$. The errorbars show the range of values from the four quasi-random arrays that were simulated.

## Conclusion

Our results demonstrate that introducing structural randomness, whether in vertical position and/or length of PrvNWs, the refractive index environment can significantly influence the optical performance of PrvNW arrays. When carefully engineered, such an out-of-plane disorder can be harnessed to enhance light absorption and improve the $J_{sc}$ (by up to 28% in our study), offering a promising pathway for the development of high-efficiency solar cells with PrvNWs in the active layer. Our results suggest strong light trapping for vertically disordered MAPI PrvNWs with as low $n_{ETL/HTL}$ as possible. Also, we showed that it is not necessary to aim for a very narrow length distribution in the nanowires, easing the fabrication requirements.

## Acknowledgements

We acknowledge the financial support from the Research Council of Finland project 359066 (HPC-Phot) and the computer resources of the Finnish IT Center for Science (CSC) and the FGCI project (Finland).

# Supporting Information

## Efficient Light Trapping Strategy in Solar Cells via Vertically Disordered Perovskite Nanowires

Sina Salimian and Nicklas Anttu

Physics, Faculty of Science and Engineering, Åbo Akademi University, FI-20500 Turku, Finland

E-mail: sina.salimian@abo.fi

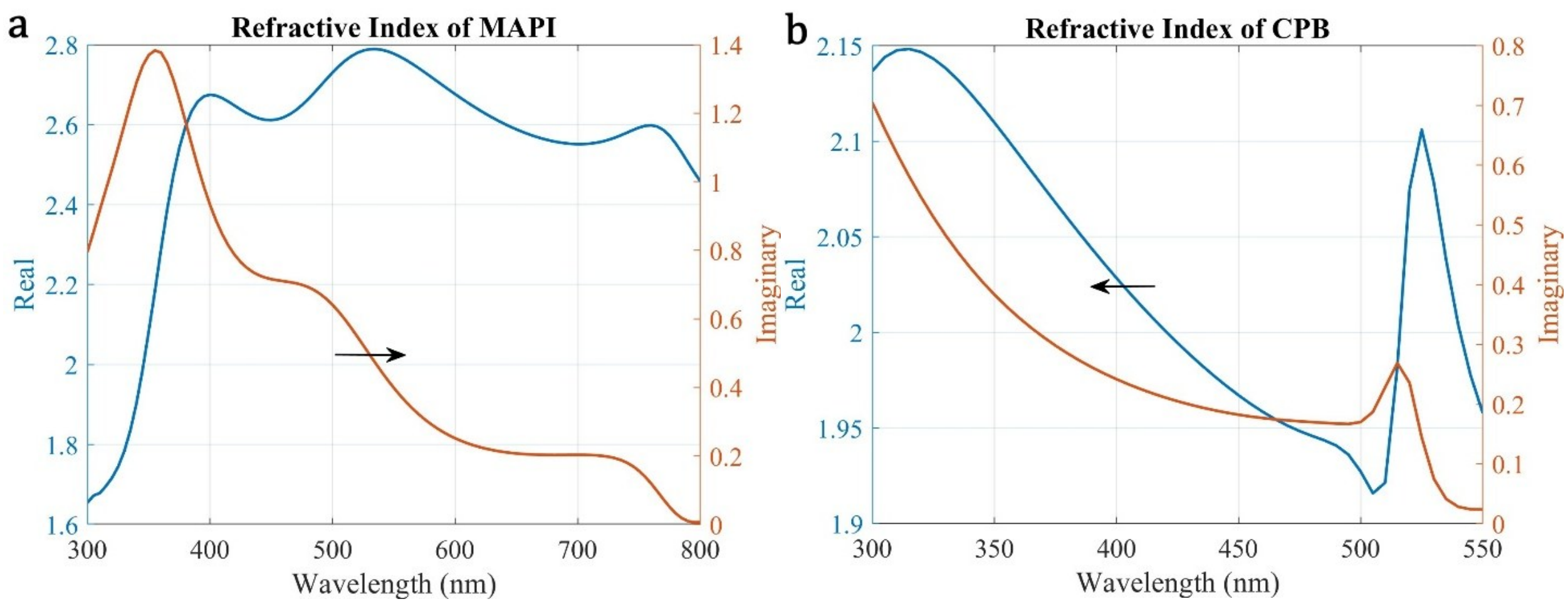

Fig. S1. Real and imaginary parts of the refractive indices of MAPI (a) and CPB (b), as extracted from Refs. [22,27].

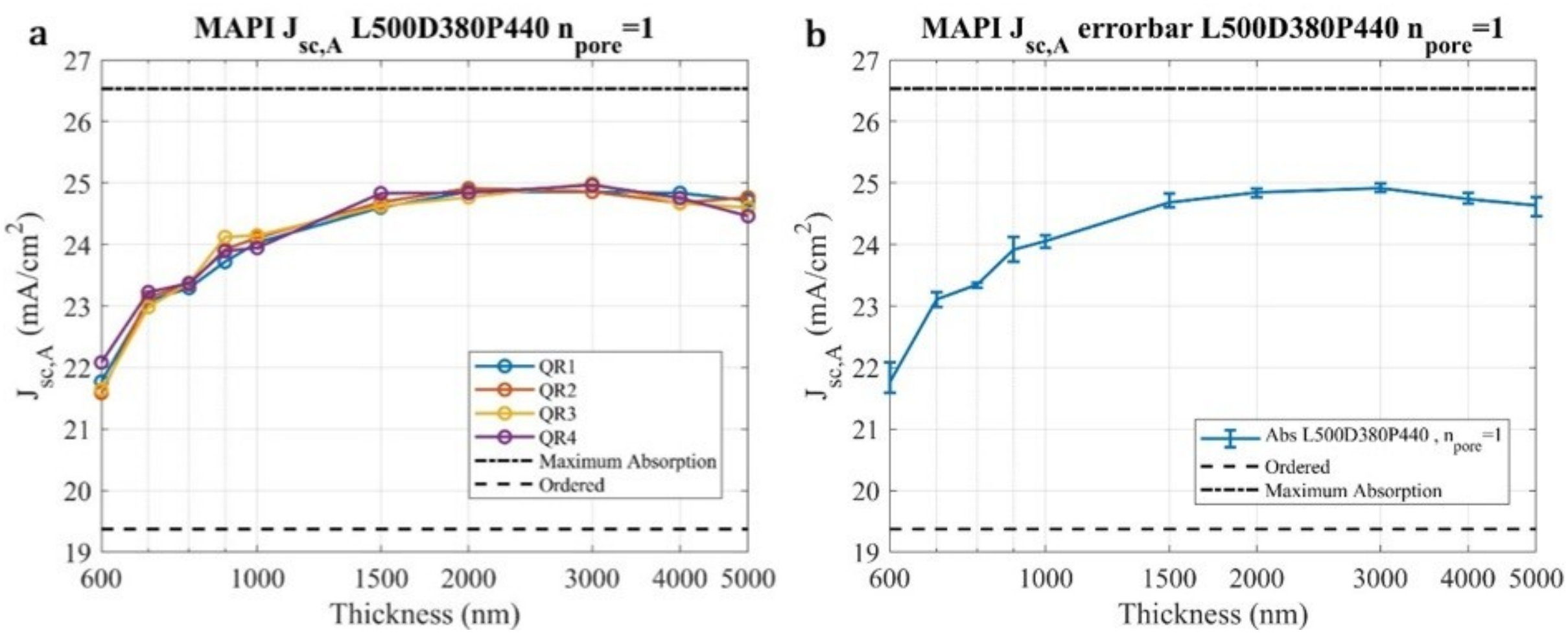

Fig. S2. For each thickness of AAO membrane, we simulated four different arrangements, that is quasi-random realizations, of PrvNWs with the same length, diameter, and period for their arrays, with the only difference in the vertical positioning of the PrvNWs in each pore of the AAO membrane. Their extracted $J_{sc}$ (a) were then taken into account with errorbars (b) to account for the extent of variation in $J_{sc}$ for different arrangements with the same dimensions.

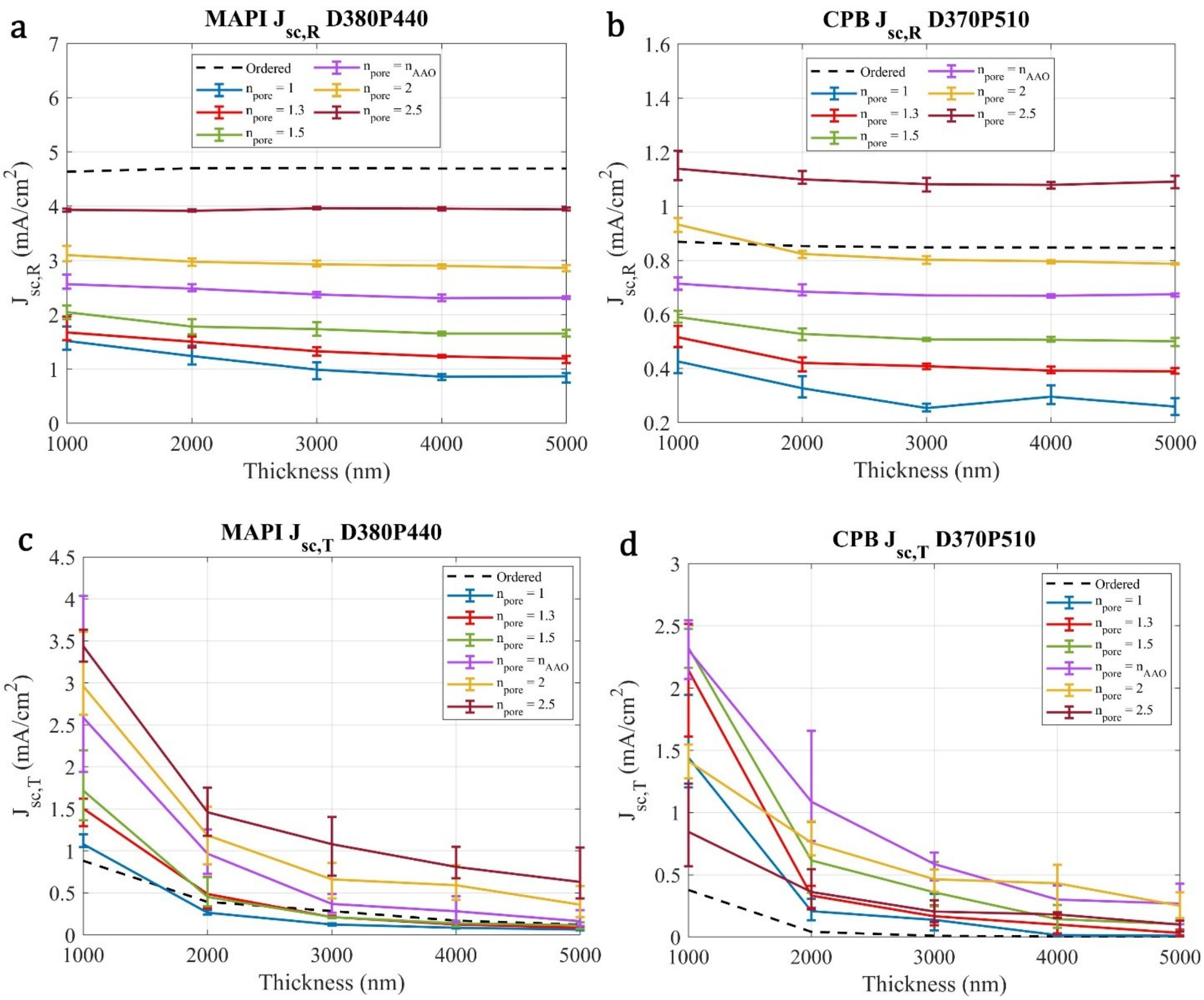

Fig. S3. Calculated $J_{sc,R}$ in (a) and (b) and $J_{sc,T}$ in (c) and (d) for the same case of varying length of the PrvNWs as for $J_{sc,A}$ in the main text in Figure 5. Here, we consider MAPI PrvNWs in (a) and (c) and CPB PrvNWs in (b) and (d). The errorbars show the range of values from the four quasi-random arrays that were simulated.